# Quantum anomalous Hall effect driven by magnetic proximity coupling in all-telluride based heterostructure

R. Watanabe[1], R. Yoshimi[2,a)], M. Kawamura[2], M. Mogi[1], A. Tsukazaki[3], X. Z. Yu[2], K. Nakajima[2], K. S. Takahashi[2,4], M. Kawasaki[1,2] and Y. Tokura[1,2,5]

[1] Department of Applied Physics and Quantum-Phase Electronics Center (QPEC), University of Tokyo, Tokyo 113-8656, Japan.

[2] RIKEN Center for Emergent Matter Science (CEMS), Wako 351-0198, Japan.

[3] Institute for Materials Research, Tohoku University, Sendai 980-8577, Japan

[4] PRESTO, Japan Science and Technology Agency (JST), Chiyoda-ku, Tokyo 102-0075, Japan.

[5]Tokyo College, University of Tokyo, Bunkyo-ku, Tokyo 113-8656, Japan.

a) Author to whom correspondence to be addressed: ryutaro.yoshimi@riken.jp

**Abstract**

The quantum anomalous Hall effect (QAHE) is an exotic quantum phenomenon originating from dissipation-less chiral channels at the sample edge. While the QAHE has been observed in magnetically doped topological insulators (TIs), exploiting magnetic proximity effect on the TI surface from adjacent ferromagnet layers may provide an alternative approach to the QAHE by opening an exchange gap with less disorder than that in the doped system. Nevertheless, the engineering of a favorable heterointerface that realizes the QAHE based on the magnetic proximity effect remains to be achieved. Here, we report on the observation of the QAHE in a proximity coupled system of non-magnetic TI and ferromagnetic insulator (FMI). We have designed sandwich heterostructures of $(Zn,Cr)Te/(Bi,Sb)_2Te_3/(Zn,Cr)Te$ that fulfills two prerequisites for the emergence of the QAHE; the formation of a sizable exchange gap at the TI surface state and the tuning of the Fermi energy into the exchange gap. The efficient proximity coupling in the all-telluride based heterostructure as demonstrated here will enable a realistic design of versatile tailor-made topological materials coupled with ferromagnetism, ferroelectricity, superconductivity, and so on.

Three-dimensional topological insulator (TI) is a novel state of matter hosting insulating bulk and conducting surface states as protected by time-reversal symmetry[1,2]. The breaking of time reversal symmetry at the surface states of a TI leads to the formation of an exchange gap and the emergence of chiral edge states, which gives the quantum anomalous Hall effect (QAHE) with the quantized Hall resistance to $h/e^2$ ($h$ is the Planck's constant and $e$ is elementary charge) when the Fermi level ($E_F$) is tuned within the exchange gap [Fig. 1(a)][3-5]. Regarding the emergence of chiral edge channels, the QAHE is phenomenologically equivalent to the integer quantum Hall effect (IQHE) which occurs under magnetic field[6], but is different in the microscopic mechanism of gap formation; the Landau level splitting by cyclotron motion of carriers for the QHE versus the magnetic exchange interaction for the QAHE. The distinct nature of the QAHE that does not need external magnetic field has unveiled unique quantum transport phenomena based on chiral edge states, such as the manipulation of edge conduction at magnetic domain walls and the detection of chiral Majorana edge modes realized by the interplay with superconductivity[7,8].

The QAHE has been studied so far in TIs doped with magnetic elements such as Cr- and V-doped (Bi,Sb)$_2$Te$_3$ (Refs: 4, 9 and 10), in which the gap formation and the $E_F$ tuning into the gap are to be simultaneously fulfilled. Magnetic proximity effect has been proposed as one other promising mechanism to induce the QAHE [Fig.1(b)][1]. When a non-magnetic TI contacts with a ferromagnetic insulator (FMI) whose magnetic moment is perpendicular to the interface, the magnetic exchange interaction via the interface can open an exchange gap at the surface state of the TI. To date, the proximity coupling has been exemplified in several FMI (*e.g.* EuS, GaN, $Y_3Fe_5O_{12}$ (YIG), and $Tm_3Fe_5O_{12}$ (TIG))/TI heterostructures with observation of the anomalous Hall effect (AHE)[11-15]. Although these heterostructures indicate an advantage in terms of a wide variety of materials choice for FMI and TI, it remains still elusive to design a preferable FMI/TI/FMI sandwich heterostructure [Fig. 1(b)] that maximizes the exchange gap at top and bottom surface states. In particular, the tangent of Hall angle ($\tan\theta_H = \sigma_{xy}/\sigma_{xx}$, the ratio of transverse to longitudinal conductivity), which is a measure of how close the Fermi level to the exchange gap, has been far below 0.01 for so-far reported FMI/TI magnetic proximity systems, while $\tan\theta_H$ tends to diverge to infinity or at least exceeds unity to reach the QAHE at a moderately low temperature, *e.g.* 1 K.

To achieve a sizable exchange gap, one of the most essential parameters is the strength of exchange coupling between electrons on the surface state of TI and the localized spins in the FMI, which should be highly materials-dependent. We consider that the combination of Te-based TI and Te-based FMI comprising 3*d*-electron transition-metal element may give a strong exchange coupling for the following reason. Since Te is incorporated in both materials in common, the topological surface states originating from the 5*p*-orbital of Te may deeply extend into the FMI. As revealed by spectroscopy measurements and first-principle calculations for magnetically doped TIs such as Cr- and V-doped (Bi,Sb)$_2$Te$_3$ (Refs: 16-18), the energy levels of spin polarized density of states for the 3$d$ magnetic elements are close to 5$p$-orbitals of Te, leading to the large exchange gap formation at the surface states[19, 20]. Thus, a strong hybridization can similarly be expected in such proximity coupled systems between $p$-orbital of Te in a TI/FMI and $d$-orbital in a FMI.

We have chosen (Bi$_{1-y}$Sb$_y$)$_2$Te$_3$ (BST) as a TI and Zn$_{1-x}$Cr$_x$Te (ZCT) as a FMI. In addition to the reasons mentioned above, the combination of BST and ZCT has the following advantages. First, it is known that the Fermi energy $E_F$ of (Bi$_{1-y}$Sb$_y$)$_2$Te$_3$ can be tuned by the Sb composition $y$ (Ref: 21) and that the surface dominant electrical transport such as integer QHE has been demonstrated[22]. Second, ZnTe, a parent compound of ZCT, is an insulator with a relatively large band gap of 2.28 eV (Ref: 23) and shows a much higher resistivity than BST. Doped with Cr, ZCT works as a FMI (with the magnetization perpendicular to the film plane[24]). Third, the in-plane lattice constant of ZnTe(111) (0.432 nm) is close to those of Sb$_2$Te$_3$ (0.426 nm) and Bi$_2$Te$_3$ (0.439 nm), which helps to form a heterostructure with smooth interfaces that facilitate the extension of the surface-state wave function into the FMI layer.

We fabricated a ZCT (10 nm)/BST (8 nm)/ZCT (10 nm) sandwich heterostructure by molecular-beam epitaxy (see Sections S1 and S2 in supplementary material (SM)). A 2-nm-thick ZnTe buffer layer was adopted to improve crystallinity of the ZCT layer. Atomic-scale structure and chemical composition of the heterostructure are analyzed by cross sectional high-angle annular dark-field scanning transmission electron microscopy [Fig. 1(c)] and energy-dispersive x-ray spectroscopy (see Fig. S2 in SM). The abrupt structural change between the BST and ZCT layers with the sharp interfaces can be seen. The topological surface states are expected to locate at around the interfaces. Diffusion of Cr into the BST layer is fairly small or at most not large enough to cause the Cr-doping induced QAHE effect as observed in an optimally Cr-doped BST film (see discussions in Section S3 in SM). We defined the Hall-bar devices with using a UV photolithography and subsequent wet etching processes for transport measurements (see Section S4 in SM). Figure 1(d) shows temperature dependence of sheet resistance $R_{xx}$ for the sandwich heterostructure and the ZCT film. Here, the Cr composition $x$ for ZCT and Sb composition $y$ for BST are set at 0.17 and 0.6, respectively. $R_{xx}$ of the sandwich heterostructure film shows weak temperature dependence with a value of around $10^4$ Ω, whereas $R_{xx}$ of the ZCT film exceeds $10^8$ Ω even at room temperature and further increases as lowering temperature. The large difference in $R_{xx}$ evidences that the electric current mainly flows in the BST layer with topological surface states. As shown in Fig. 1(e), the magnetization curve for the ZCT (shown in gray) at $T = 2$ K shows a clear hysteresis, representing the ferromagnetism in the ZCT film. Magnetic-field dependence of anomalous Hall resistance $R_{yx}^{AHE}$ well agrees with that of $M$ in the ZCT film, showing similar coercive fields (magnetization curve for other temperature is shown in Fig. S6). Here $R_{yx}^{AHE}$ is defined by subtracting ordinary Hall component $R_{yx}^{AHE} = R_{yx} - R_0 B$, where $R_0$ and $B$ are ordinary Hall coefficient and magnetic field, respectively. Furthermore, the ferromagnetic transition temperatures evaluated from $M$ and $R_{yx}$ are 60 K and 40 K, respectively, which are close with each other (see Section S5 in SM). This agreement indicates that the AHE is induced by the magnetic proximity coupling with $M$ in the ZCT layers.

Figure 2(a) displays the temperature dependence of $R_{xx}$ and $R_{yx}$ for the same sandwich heterostructure under $B = 2$ T where the ZCT magnetization is saturated. $R_{yx}$ shows monotonic increase with decreasing $T$ from the onset temperature of 50 K. As $T$ is lowered below 0.1 K, $R_{yx}$ reaches the quantum resistance $h/e^2$ (~ 25.8 kΩ) while $R_{xx}$ approaches zero, showing the QAHE (the same data in the conductance form is shown in Fig. S9 in SM). A similar temperature dependence is found in the Cr-doped BST (Ref: 25) that also shows the QAHE at below 0.1 K (Fig. S10 in SM), suggesting that the spatial inhomogeneity of the exchange gap also occurs more or less in the present sandwich heterostructure due to random Cr distribution in the FMI layer and may similarly hinder the system from localization at higher temperatures. In the magnetic field dependence at the lowest temperature $T = 0.03$ K [Fig. 2(b)], $R_{yx}$ is nearly constant against $B$ at $R_{yx} = \pm h/e^2$ in the magnetic field range $|B| > 0.8$ T where the magnetization of ZCT is saturated. The observed $R_{yx}$ originates from the AHE induced by the interaction between itinerant electrons and localized magnetic moments, excluding possible contributions of the ordinary Hall effect induced by electron cyclotron motion. In addition, the $R_{xx}$ peaks in the magnetic field dependence [Fig. 2(b)] reflect the magnetic multi-domain structure of the ZCT layer during the magnetization reversal process. This means that the edge current can be controlled by $M$ in adjacent FMI layers, which is different from magnetically-doped TI systems. These observations constitute an evidence for the QAHE induced by the magnetic proximity effect in the ZCT/BST/ZCT heterostructure system.

Next, we examine the $E_F$ position dependence of the AHE by changing Sb composition $y$ in the BST channel layer. The Cr compositions for top and bottom ZCT layers are set to be the same, fixed at $x = 0.17$ in this series. Figures 3(a), 3(b), and 3(c) display $B$ dependence of $R_{yx}$ for the samples with $y = 0.50$, 0.60 and 0.65, respectively. Schematic band structure with the anticipated $E_F$ position is shown on the top of each panel. The spontaneous Hall resistance is the largest at $y = 0.60$ as shown in Figs. 3(a)-3(c), indicating the $E_F$ of $y = 0.60$ is closely tuned to the exchange gap where the Berry curvature is maximized. This trend is further supported by the sign change of ordinary Hall coefficient; negative slope of $R_{yx}$ for $y = 0.50$ [Fig. 3(a)] and 0.60 [Fig. 3(b)] and positive for $y = 0.65$ [Fig. 3(c)]. The dominant carrier-type is converted from electron to hole across the charge neutral point (CNP) in between $y = 0.60$ and 0.65. Figures 3(d) and 3(e) respectively summarizes the $y$ dependence of the carrier density/type and the anomalous Hall resistance $R_{yx}^{AHE}$ (the $R_{yx}$-$B$ data for all the samples are presented in Fig. S11(a) in SM). Systematic variation of carrier density/type ensures that the $E_F$ position is well regulated by the Sb composition. Although the optimum $y$ at around 0.6 is slightly shifted from $y$ = 0.85-0.95 in a single-layer BST[21,22], the sharp and sizable peak in $R_{yx}^{AHE}$ (Fig. 3(e)) shows up around the CNP, which is in accord with the fact that the QAHE is observed at low temperatures below 0.1 K.

Finally, the relationship between ferromagnetic $T_C$ of the FMI layer and the AHE is discussed with a measure of the tangent of Hall angle, $\tan\theta_H = \sigma_{yx} / \sigma_{xx}$ defined under saturated magnetization at $B$ = 2 T. Figure 4(a) shows temperature dependence of $\tan\theta_H$ for ZCT/BST/ZCT sandwich heterostructures with different Cr compositions $x$ (temperature dependence of $R_{xx}$ and $R_{yx}$ for samples with different $y$ are shown in Fig. S10 in SM). Sb composition for the BST layer is tuned at the optimal value $y$ = 0.60 in this series. For the samples with $x \leq 0.17$, $\tan\theta_H$ monotonically increases with decreasing $T$ below the critical temperature. The maximum of the $\tan\theta_H$ for $x$ = 0.17 exceeds 2.5 at $T$ = 0.5 K, manifesting that the system is approaching the QAH state. In contrast, the samples with $x$ > 0.17 have very small $\tan\theta_H$ over a wide temperature range. Figure 4(b) displays a color contour plot of $\tan\theta_H$ as functions of $x$ and $T$ together with the $x$-dependence of ferromagnetic transition temperature (black symbols), the latter of which was estimated from Arrott plot analysis of anomalous Hall resistance (Fig. S7 in SM) and referred as $T_C^*$. Despite the continuous increase in $T_C^*$ up to $x = 0.35$, $\tan\theta_H$ turns to decrease above $x = 0.17$. This is probably due to the segregation of CrTe in the ZCT layers (see Section S6 and Fig. S12 in SM), resulting in the reduction of resistivity of ZCT layer[26,27] or degradation of crystal quality of the BST layer. As far as the crystal structure of the FMI layer is maintained in $x \leq 0.17$, the observable temperature of the QAHE increases with the composition of the magnetic element, which means that a suitable FMI with higher $T_C$ could increase the observable temperature of the QAHE, *e.g.* by donor doping to ZCT (Ref: 28).

To summarize, we have observed the QAHE driven by magnetic proximity coupling in $Zn_{1-x}Cr_xTe$ (ZCT) /$(Bi_{1-y}Sb_y)_2Te_3$(BST)/ZCT sandwich heterostructures. The observed anomalous Hall response faithfully reflects the magnetic properties of the FMI layer, ensuring the magnetic proximity effect. Clear signatures of the QAHE with quantized $R_{yx}$ and vanishing $R_{xx}$ are observed, when the precise tuning of $E_F$ and the relatively high $T_C$ in the ferromagnetic ZCT layer are attained. The key strategy to design a heterostructure is the strong exchange coupling realized through the interface between the all-telluride based TI and FMI. It is noteworthy that the complex telluride materials involve families of not only these TIs and ferromagnets, but also ferroelectrics[29,30] and superconductors[31,32]. The present work would pave a way for the exploration of all-telluride based heterostructures that would realize even more exotic topological quantum phenomena.

**Acknowledgements** This research was supported by the Japan Society for the Promotion of Science through JSPS/MEXT Grant-in-Aid for Scientific Research (No. 15H05853, No. 15H05867, No. 17H04846, No. 18H04229 and No. 18H01155), and JST CREST (No. JPMJCR16F1).

**Reference**

1. X.-L. Qi, T. L. Hughes, S. -C. Zhang, *Phys. Rev. B* **78**, 195424 (2008).

2. M. Z. Hasan, C. L. Kane, *Rev. Mod. Phys.* **82**, 3045–3067 (2010).

3. R. Yu, W. Zhang, H.-J. Zhang, S.-C. Zhang, X. Dai, Z. Fang, *Science* **329**, 61-64 (2010).

4. C.-Z. Chang, J. Zhang, X. Feng, J. Shen, Z. Zhang, M. Guo, K. Li, Y. Ou, P. Wei, L.-L. Wang, Z.-Q. Ji, Y. Feng, S. Ji, X. Chen, J. Jia, X. Dai, Z. Fang, S.-C. Zhang, K. He, Y. Wang, L. Lu, X.-C. Ma, Q.-K. Xue, *Science* **340**, 167-170 (2013).

5. Y. Tokura, K. Yasuda, A. Tsukazaki, *Nat. Rev. Phys.* **1,** 126-143 (2019).

6. K. v. Klitzing, G. Dorda, M. Pepper, *Phys. Rev. Lett.* **45**, 494-497 (1980).

7. K. Yasuda, M. Mogi, R. Yoshimi, A. Tsukazaki, K. S. Takahashi, M. Kawasaki, F. Kagawa, Y. Tokura, *Science* **358**, 1311-1314 (2017).

8. Q. L. He, L. Pan, A. L. Stern, E. C. Burks, X. Che, G. Yin, J. Wang, B. Lian, Q. Zhou, E. S. Choi, K. Murata, X. Kou, Z. Chen, T. Nie, Q. Shao, Y. Fan, S.-C. Zhang, K. Liu, J. Xia, K. L. Wang., *Science* **357**, 294-299 (2017).

9. J. G. Checkelsky, R. Yoshimi, A. Tsukazaki, K. S. Takahashi, Y. Kozuka, J. Falson, M. Kawasaki,

Y. Tokura, *Nat. Phys.* **10**, 731-736 (2014).

10. C.–Z. Chang, W. Zhao, D. Y. Kim, H. Zhang, B. A. Assaf, D. Heiman, S.-C. Zhang, C. Liu, M. H. W. Chan, J. S. Moodera, *Nat. Mater.* **14**, 473-477 (2015).

11. Q. I. Yang, M. Dolev, L. Zhang, J. Zhao, A. D. Fried, E. Schemm, M. Liu, A. Palevski, A. F. Marshall, S. H. Risbud, A. Kapitulnik, *Phys. Rev. B* **88**, 081407(R) (2013).

12. F. Katmis, V. Lauter, F. S. Nogueira, B. A. Assaf, M. E. Jamer, P. Wei, B. Satpati, J. W. Freeland, I. Eremin, D. Heiman, P. Jarillo-Herrero, J. S. Moodera, *Nature* **533**, 513-516 (2016).

13. A. Kandala, A. Richardella, D. W. Rench, D. M. Zhanga, T. C. Flanagan, and N. Samarth, *Appl. Phys. Lett.* **103**, 202409 (2013).

14. Z. Jiang, C.-Z. Chang, C. Tang, P. Wei, J. S. Moodera, J. Shi, *Nano Lett.* **15**, 5835-5840 (2015).

15. C. Tang, C.-Z. Chang, G. Zhao, Y. Liu, Z. Jiang, C.-X. Liu, M. R. McCartney, D. J. Smith, T. Chen, J. S. Moodera, J. Shi, *Sci. Adv.* **3**, e1700307 (2017).

16. M. Ye, W. Li, S. Zhu, Y. Takeda, Y. Saitoh, J. W., H. Pan, M. Nurmamat, K. Sumida, F. Ji, Z. Liu, H. Yang, Z. Liu, D. Shen, A. Kimura, S. Qiao, X. Xie, *Nat. Commun.* **6**, 8913 (2015).

17. G. Vergniory, M. M. Otrokov, D. Thonig, M. Hoffmann, I. V. Maznichenko, M. Geilhufe, X. Zubizarreta, S. Ostanin, A. Marmodoro, J. Henk, W. Hergert, I. Mertig, E. V. Chulkov, A. Ernst, *Phys. Rev. B* **89**, 165202 (2014).

18. T. R. F. Peixoto, H. Bentmann, S. Schreyeck, M. Winnerlein, C. Seibel, H. Maaß, M. Al-Baidhani, K. Treiber, S. Schatz, S. Grauer, C. Gould, K. Brunner, A. Ernst, L. W. Molenkamp, F. Reinert, *Phys. Rev. B* **94,** 195140 (2016).

19. M. M. Otrokov, T. V. Menshchikova, M. G. Vergniory, I. P. Rusinov, A. Y. Vyazovskaya, Y. M. Koroteev, G. Bihlmayer, A. Ernst, P. M. Echenique, A. Arnau, *2D Mater.* **4,** 025082 (2017).

20. S. V. Eremeev, V. N. Men'shov, V. V. Tugushev, P. M. Echenique, E. V. Chulkov, *Phys. Rev. B*

**88,** 144430 (2013).

21. J. Zhang, C.-Z. Chang, Z. Zhang, J. Wen, X. Feng, K. Li, M. Liu, K. He, L. Wang, X. Chen, Q.-K. Xue, X. Ma, Y. Wang, *Nat. Commun.* **2,** 574 (2011).

22. R. Yoshimi, A. Tsukazaki, Y. Kozuka, J. Falson, K. S. Takahashi, J. G. Checkelsky, N. Nagaosa, M. Kawasaki, Y. Tokura, *Nat. Commun.* **6**, 6627 (2015).

23. K. Sato, S. Adachi, *J. Appl. Phys.* **73**, 926-931 (1993).

24. H. Saito, W. Zaets, S. Yamagata, Y. Suzuki, K. Ando, *J. Appl. Phys.* **91**, 8085-8087 (2002).

25. M. Mogi, R. Yoshimi, A. Tsukazaki, K. Yasuda, Y. Kozuka, K. S. Takahashi, M. Kawasaki, Y. Tokura, *Appl. Phys. Lett.* **107,** 182401 (2015).

26. S. Polesya *et*, S. Mankovsky, D. Benea, H. Ebert and W. Benschal, *J. Phys.: Condens. Matter* **22**,156002 (2010).

27. D. Zhao, L. Zhang, I. A. Malik, M.-G. Liao, W.-Q. Cui, X.-Q. Cai, C. Zheng, L. Li, X.-P. Hu, D. Zhang, J.-X. Zhang, X. Chen, W.-J. Jiang, Q.-K. Xue, *Nano Research* **11,** 3116 (2018).

28. N. Ozaki, N. Nishizawa, S. Marcet, S. Kuroda, O. Eryu, K. Takita, *Phys. Rev. Lett.* **97,** 037201 (2006).

29. D. D. Sante, P. Barone, R. Bertacco, S. Picozzi, *Adv. Mater.* **25,** 509-513 (2012).

30. R. Yoshimi, K. Yasuda, A. Tsukazaki, K. S. Takahashi, M. Kawasaki, Y. Tokura, *Sci. Adv.* **4,** eaat9989 (2018).

31. T. Hanaguri, S. Niitaka, K. Kuroki, H. Takagi, *Science* **328,** 474-476 (2010).

32. A. S. Erickson, J. –H. Chu, M. F. Toney, T. H. Geballe, I. R. Fisher, *Phys. Rev. B* **79,** 024520 (2009).

**Figure Legends**

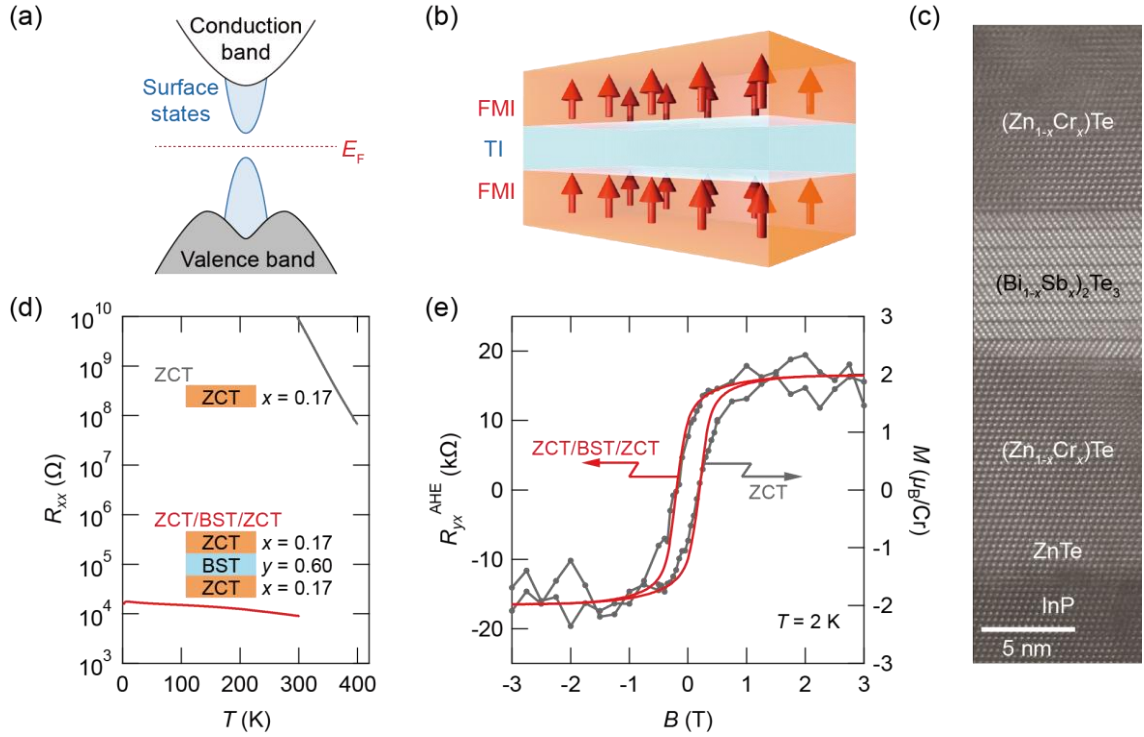

FIG. 1 (a) Schematic band structure for a TI with a gapped surface state. $E_F$ represents the Fermi energy. (b) Schematic drawing of a FMI/TI/FMI sandwich heterostructure. (c) A cross sectional high-angle annular dark-field scanning transmission electron microscopy (HAADF-STEM) image of a $Zn_{1-x}Cr_xTe$(ZCT)/$(Bi_ySb_{1-y})_2Te_3$(BST)/ZCT sandwich heterostructure, here $x = 0.17$ and $y = 0.60$. The scale bar is 5 nm. (d) Temperature dependence of sheet resistance $R_{xx}$ for a 10-nm-thick ZCT single-layer film ($x = 0.17$) and the ZCT/BST/ZCT ($x = 0.17$, $y = 0.60$) sandwich heterostructure. (e) Magnetic field dependence of anomalous Hall resistance $R_{yx}^{AHE}$ of the ZCT/BST/ZCT ($x = 0.17$, $y = 0.60$) sandwich heterostructure (red) and magnetization $M$ of 10-nm thick ZCT thin film (gray) at $T = 2$ K.

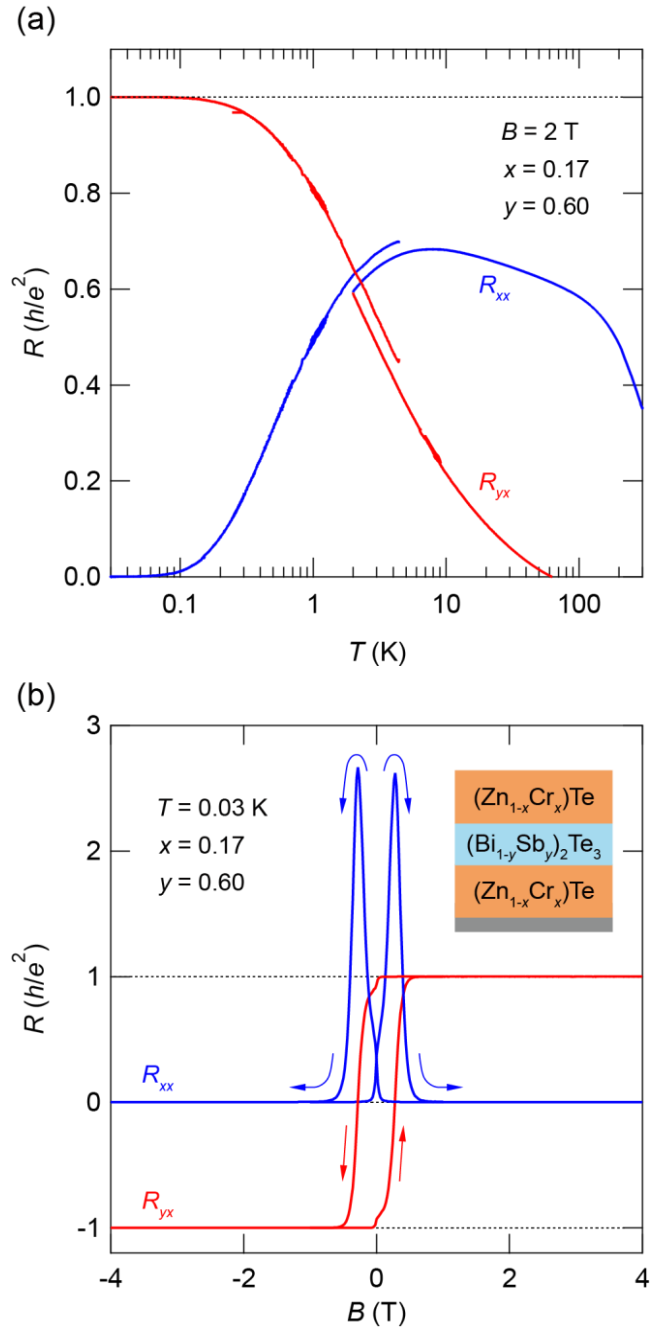

FIG. 2 (a) Temperature dependence of $R_{xx}$ (blue) and $R_{yx}$ (red) under magnetic field $B = 2$ T for a ZCT/BST/ZCT heterostructure with $x = 0.17$ and $y = 0.60$. (b) Magnetic field dependence of $R_{xx}$ (blue) and $R_{yx}$ (red) at $T = 0.03$ K for the same sample. The inset shows a schematic of the heterostructure.

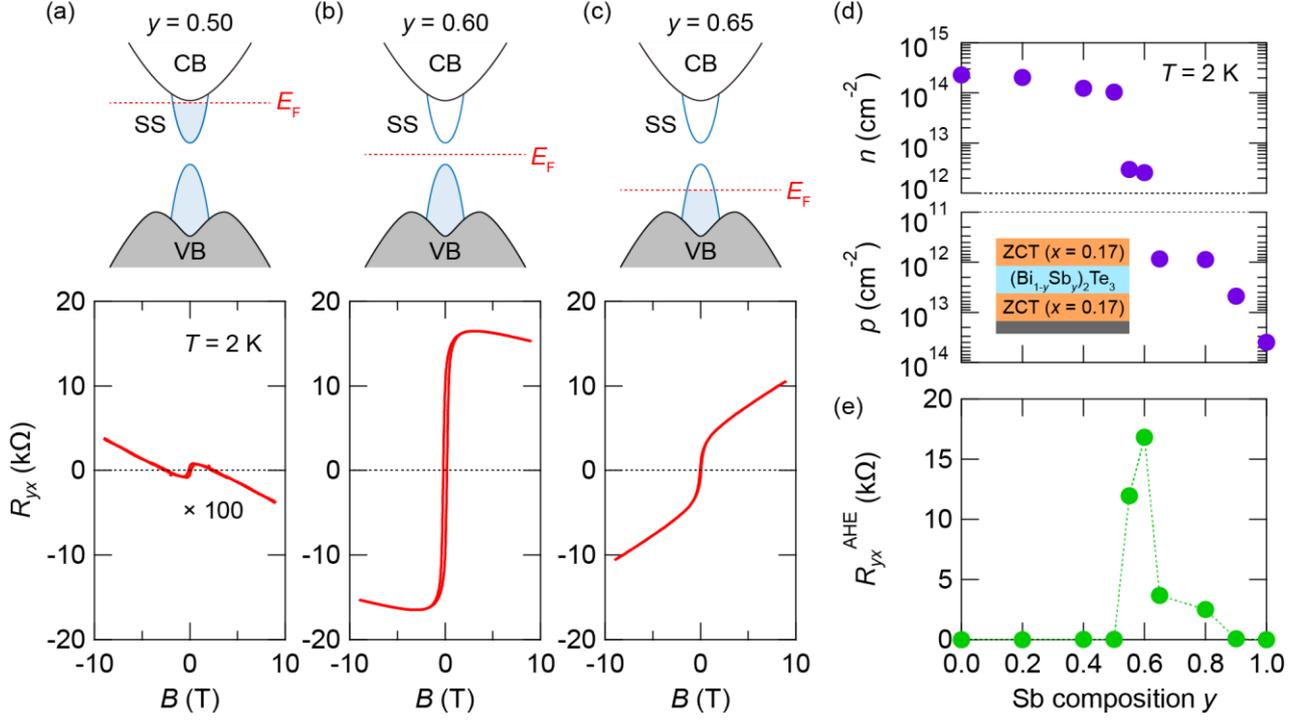

FIG. 3 (a)-(c) Top: Schematic band structures of BST of the ZCT/BST/ZCT sandwich heterostructure. CB, VB and SS denote the conduction band, valence band and surface states, respectively. Bottom: Magnetic field dependence of Hall resistance $R_{yx}$ at $T = 2$ K. (a), (b) and (c) are for Sb composition $y = 0.50$, $0.60$ and $0.65$, respectively. (d), (e) Sb composition $y$ dependence of sheet carrier concentration at $T = 2$ K evaluated from the ordinary Hall component for $B > 5$ T (d) and anomalous Hall resistance $R_{yx}^{AHE}$ at $B = 2$ T (e). The inset of (d) shows a schematic of the heterostructure.

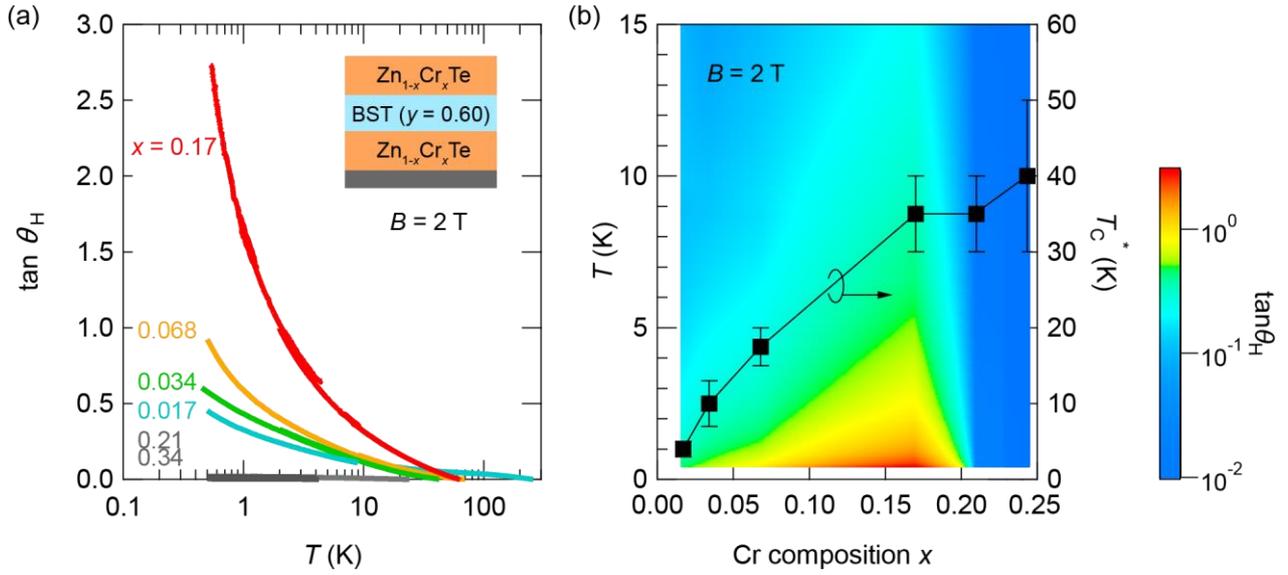

FIG. 4 (a) Temperature dependence of the tangent of Hall angle $\tan\theta_H$ under magnetic field of $B = 2$ T for ZCT/BST/ZCT sandwich heterostructures with various Cr composition $x$. The inset shows a schematic of the heterostructure. (b) Color contour plot of $\tan\theta_H$ in the plane of temperature $T$ and Cr composition $x$. The ferromagnetic transition temperature $T_C^*$ estimated from AHE is also plotted as a function of $x$ (black closed squares).